\documentclass[runningheads]{llncs}

\usepackage[T1]{fontenc}
\usepackage{graphicx}
\usepackage{listings}
\usepackage{graphicx}
\graphicspath{{figures}}
\usepackage{makecell}
\usepackage{bm}
\usepackage{multirow}
\usepackage{footmisc}
\usepackage[table]{xcolor}
\usepackage{hhline}
\usepackage{wrapfig}
\usepackage{array}

\newcounter{mysfig}
\counterwithin{mysfig}{figure}

\renewcommand\themysfig{(\alph{mysfig})}
\makeatletter
\newcommand\Scaption[1]{%
\refstepcounter{mysfig}%
\vskip.5\abovecaptionskip
  \sbox\@tempboxa{\small\themysfig~#1}%
  \ifdim \wd\@tempboxa >\hsize
    \small\themysfig~#1\par
  \else
    \global \@minipagefalse
    \hb@xt@\hsize{\hfil\box\@tempboxa\hfil}%
  \fi
  \vskip\belowcaptionskip}
\makeatother

\newlength{\Oldarrayrulewidth}

\newcolumntype{C}[1]{>{\centering}m{#1}}

\begin{document}

\title{Leveraging Deep Reinforcement Learning for Metacognitive Interventions across \\ Intelligent Tutoring Systems}


\titlerunning{Deep Reinforcement Learning for Metacognitive Interventions}
%


\author{Mark Abdelshiheed \and John Wesley Hostetter \and \\Tiffany Barnes \and Min Chi\\ \{mnabdels,\, jwhostet,\, tmbarnes,\, mchi\}@ncsu.edu}

 \authorrunning{Abdelshiheed et al.}

 \institute{North Carolina State University, Raleigh, NC 27695, USA}

\maketitle              
\begin{abstract}

This work compares two approaches to provide metacognitive interventions and their impact on preparing students for future learning across Intelligent Tutoring Systems (ITSs). In two consecutive semesters, we conducted two classroom experiments: Exp. 1 used a classic artificial intelligence approach to \textit{classify} students into different metacognitive groups and provide \textit{static} interventions based on their classified groups. In Exp. 2, we leveraged Deep Reinforcement Learning (\textit{DRL}) to provide \textit{adaptive} interventions that consider the dynamic changes in the student's metacognitive levels. In both experiments, students received these interventions that taught \textit{how} and \textit{when} to use a backward-chaining (BC) strategy on a logic tutor that supports a default forward-chaining strategy. Six weeks later, we trained students on a probability tutor that only supports BC without interventions. Our results show that adaptive DRL-based interventions closed the metacognitive skills gap between students. In contrast, static classifier-based interventions only benefited a subset of students who knew \textit{how} to use BC in advance. Additionally, our DRL agent prepared the experimental students for future learning by significantly surpassing their control peers on both ITSs.

\keywords{Reinforcement Learning \and Artificial Intelligence \and Intelligent Tutoring Systems \and Metacognitive Interventions \and Metacognitive Skills }
\end{abstract}

\section{Introduction}

A challenging desired aspect of learning is being continuously prepared for future learning \cite{bransford1999transferRethinking}. Our incremental knowledge is the evidence that preparation for future learning exists yet is hard to predict and measure \cite{detterman1993transfer}. Considerable research has found that one factor that facilitates preparing students for future learning is their metacognitive skills \cite{belenky2009metacogTransfer,zepeda2015INSTRUCTION}. We focus on two types of metacognitive skills related to problem-solving strategies: \emph{strategy-awareness} \cite{abdelshiheed2021preparing,roberts1993metacogDefinitionStrategySelection2,shabrina2023investigating} and \emph{time-awareness} \cite{abdelshiheed2020metacognition,de2018longSWITCH-INSTRUCTION}, that are respectively, \emph{how} and \emph{when} to use each strategy.

Substantial work has demonstrated that metacognitive interventions of domain knowledge or problem-solving strategies accelerate preparation for future learning \cite{belenky2009metacogTransfer,richey2015promptsWE,shabrina2023investigating} and promote strategy- and time-awareness \cite{fellman2020Presented,sporer2009Presented,zepeda2015INSTRUCTION}. Such interventions included hints, feedback, prompted nudges, worked examples, and direct strategy presentation. 
However, these interventions were \emph{static or hard-coded} into the learning environment despite the fact that students often acquire and master metacognitive skills as they learn
\cite{kuhn2000metacognitiveDevelopement}. Reinforcement Learning (RL) \cite{sutton2018reinforcement} is one of the most effective approaches for providing adaptive support and scaffolding across Intelligent Tutoring Systems (ITSs) \cite{hostetter2023self,krueger2017MetacognitiveRL,zhou2019hierarchical}. The deep learning extension of RL, known as Deep RL (DRL), has been commonly utilized in pedagogical policy induction across ITSs \cite{ju2021CriticalPYR,sanz2020exploring} due to its higher support of model sophistication. As far as we know,  no prior work has leveraged DRL to provide \emph{adaptive} metacognitive interventions and investigated their impact on preparation for future learning across ITSs.

In this work, we conducted two consecutive experiments to compare two approaches for providing metacognitive interventions and their impact on preparing students for future learning on ITSs. In Exp. 1, we utilized a Random Forest Classifier (RFC) to classify students into different metacognitive groups and then provide static interventions based on their classification, while in Exp. 2, we leveraged a DRL-based approach for adaptive interventions that consider the dynamic changes in the student’s metacognitive levels. Based on strategy- and time-awareness, our prior work classified students into those who are \emph{both} strategy- and time-aware (\textit{StrTime}), those who are \textit{only} strategy-aware (\textit{StrOnly}), and the rest who follow the default strategy (\textit{Default}) \cite{abdelshiheed2021preparing,abdelshiheed2020metacognition}. We found that only \textit{StrTime} students were prepared for future learning, as they learned significantly better than their peers across different deductive domains. Motivated by such findings, we designed metacognitive interventions to teach students \textit{how} and \textit{when} to use a backward-chaining (BC) strategy on a logic tutor that supports a default forward-chaining strategy. After six weeks, students were trained on a probability tutor that only supports BC without receiving interventions. Our results showed that \textit{Default} and \textit{StrOnly} students benefited equally from our DRL policy and surprisingly outperformed \textit{StrTime} students. However, the RFC-based approach only helped \textit{StrOnly} students to catch up with \textit{StrTime}.

\section{Background and Related Work}

\subsection{Metacognitive Interventions for Strategy Instruction}

Metacognition indicates one's awareness of their cognition and the ability to control and regulate it \cite{flavell1979metacognitionDefinition}. Strategy- and time-awareness are two metacognitive skills that address \emph{how} and \emph{when} to use a problem-solving strategy, respectively \cite{abdelshiheed2020metacognition,de2018longSWITCH-INSTRUCTION}. Much prior work has emphasized the role of strategy awareness in preparation for future learning \cite{abdelshiheed2021preparing,shabrina2023investigating} and the impact of time awareness on academic performance and planning skills  \cite{de2018longSWITCH-INSTRUCTION,fazio2016timeAwareness}.

Considerable research has shown that metacognitive interventions promote strategy- and time-awareness \cite{fellman2020Presented,sporer2009Presented,zepeda2015INSTRUCTION}. We focus on two metacognitive interventions: directly presenting the strategy \cite{fellman2020Presented,sporer2009Presented} and prompting nudges to use it \cite{belenky2009metacogTransfer,richey2015promptsWE,zepeda2015INSTRUCTION}. Sp{\"o}rer et al. \cite{sporer2009Presented} found that students who were explicitly instructed on comprehensive reading strategies surpassed their peers, who were taught by the instructors' text interactions, on a transfer task and follow-up test. They also understood how, when, and why to use each reading strategy.

Zepeda et al. \cite{zepeda2015INSTRUCTION} demonstrated that metacognitive interventions impact learning outcomes, strategy mastery, and preparation for future learning. The experimental condition who received tutoring nudges and worked examples performed significantly better on a physics test than their control peers. They also made better metacognitive judgments and demonstrated mastery of knowing how and when to use physics strategies. As an example of preparation for future learning, the experimental students performed better on a novel self-guided `control of variables' learning task than their control peers.

Despite much prior work on metacognitive interventions, the interventions were either not adaptive, not applied to ITSs, or had no preparation for future learning assessment. In our work, we conducted two experiments to provide metacognitive interventions and investigated their impact on preparing students for future learning across ITSs. Specifically, we first attempted static metacognitive interventions using a classifier-based approach, then compared it against a DRL-based approach for adaptive metacognitive interventions.

\subsection{Reinforcement Learning in Intelligent Tutoring Systems}

Reinforcement Learning (RL) is a popular machine learning branch ideal in environments where actions result in numeric rewards without knowing a ground truth \cite{sutton2018reinforcement}. Due to its aim of maximizing the cumulative reward, RL has been widely used in educational domains due to the flexible implementation of reward functions \cite{hostetter2023self,ju2021CriticalPYR,sanz2020exploring}. Deep RL (DRL) is a field that combines RL algorithms with neural networks; for instance, Deep Q-Learning is the neural network extension of Q-Learning, where a neural network is used to approximate the Q-function \cite{mnih2015DQN}. Substantial work has used RL and DRL in inducing pedagogical policies across ITSs \cite{ju2021CriticalPYR,sanz2020exploring,zhou2019hierarchical}. Zhou et al. \cite{zhou2019hierarchical} utilized hierarchical RL to improve the learning gain on an ITS. They showed that their policy significantly outperformed an expert and a random condition.

Ju et al. \cite{ju2021CriticalPYR} presented a DRL framework that identifies the critical decisions to induce a critical policy on an ITS. They evaluated their critical-DRL framework based on two success criteria: \textit{necessity} and \textit{sufficiency}. The former required offering help in \textbf{all} critical states, and the latter required offering help \textbf{only} in critical states. Their results showed that the framework fulfilled both criteria. Sanz-Ausin et al. \cite{sanz2020exploring} conducted two consecutive classroom studies where DRL was applied to decide whether the student or tutor should solve the following problem. They found that the DRL policy with simple explanations significantly improved students' learning performance more than an expert policy.

Despite the wide use of RL and DRL on ITSs, the attempts to combine either with metacognitive learning have been minimal \cite{krueger2017MetacognitiveRL}.  Krueger et al. \cite{krueger2017MetacognitiveRL} used RL to teach the metacognitive skill of knowing how much to plan ahead (Deciding How to Decide). Their metacognitive reinforcement learning framework builds on the semi-gradient SARSA algorithm \cite{sutton2018reinforcement} that was developed to approximate Markov decision processes. They defined a meta Q-function, known as $Q_{meta}$, that takes the meta state of the environment and the planning horizon action. They evaluated their framework on two planning tasks, where the authors defined constrained reward functions, and the rewards could be predicted many steps ahead to facilitate forming a plan. 

In our work, we induced and deployed a DRL policy of metacognitive interventions on a logic tutor and investigated its impact on preparing students for future learning on a subsequent probability tutor. Additionally, we did not override any DRL mathematical definition, such as the Q-function. Instead, our DRL algorithm's metacognitive aspect resides in our interventions' nature.

\section{Logic and Probability Tutors} \label{tutors}

\vskip -0.3in
\begin{figure}[ht!]
\begin{minipage}[t]{0.38\columnwidth}
\begin{minipage}[b]{1\columnwidth}
\includegraphics[width=1\linewidth]{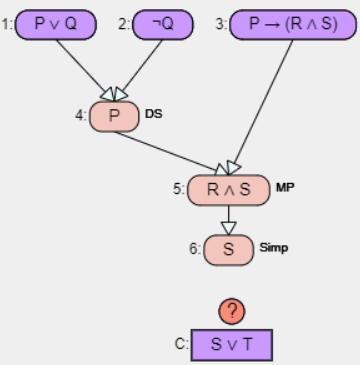}\par
\end{minipage}
\Scaption{Forward Chaining}
\label{fig:fc}
\end{minipage}\hfill
\begin{minipage}[t]{0.54\columnwidth}
\begin{minipage}[b]{\columnwidth}
\includegraphics[width=\linewidth]{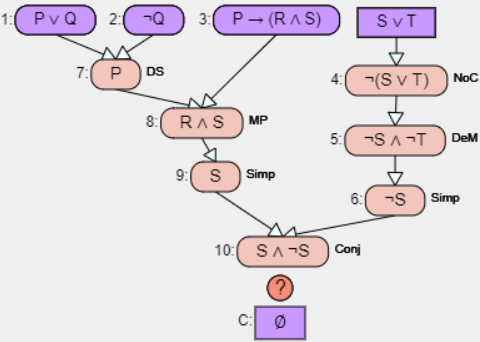}
\end{minipage}
\Scaption{Backward Chaining}
\label{fig:bc}
\end{minipage}\hfill
\caption{Logic Tutor Problem-Solving Strategies} 
\label{fig:logic}
\end{figure}

\noindent{\textbf{Logic Tutor:}} It teaches propositional logic proofs by applying valid inference rules such as Modus Ponens through the standard sequence of pre-test, training and post-test. The three phases share the same interface, but training is the \emph{only} one where students can seek and get help. The pre-test has two problems, while the post-test is harder and has six problems; the first two are isomorphic to the pre-test problems. Training consists of five ordered levels with an \emph{incremental degree of difficulty}, and each level consists of four problems. Every problem has a score in the $[0,100]$ range based on the accuracy, time and solution length.

The \emph{pre-} and \emph{post-test} scores are calculated by averaging their pre- and post-test problem scores. A student can solve any problem throughout the tutor by either a \textit{forward-chaining} or a \textit{backward-chaining (BC)} strategy. Figure \ref{fig:logic}a shows that for \emph{forward chaining}, one must derive the conclusion at the bottom from givens at the top, while Figure \ref{fig:logic}b shows that for \emph{BC}, students need to derive a contradiction from givens and the \emph{negation} of the conclusion. Problems are presented by \emph{default} in forward chaining, but students can switch to BC by clicking a button in the tutor interface.

\vskip 0.1 in
\noindent{\textbf{Probability Tutor:}} It teaches how to solve probability problems using ten principles, such as the Complement Theorem. The tutor consists of a textbook, pre-test, training, and post-test. Like the logic tutor, training is the only section for students to receive and ask for hints, and the post-test is harder than the pre-test. The textbook introduces the domain principles, while training consists of $12$ problems, each of which can \emph{only} be solved by \textit{BC} as it requires deriving an answer by \emph{writing and solving equations} until the target is ultimately reduced to the givens. 

In pre- and post-test, students solve $14$ and $20$ open-ended problems, where each pre-test problem has an isomorphic post-test problem. The answers are graded in a double-blind manner by experienced graders using a partial-credit rubric, where grades are based \emph{only} on accuracy in the $[0,100]$ range. The \emph{pre-} and \emph{post-test} scores are the average grades in their respective sections.

\vskip -0.3 in
\section{Methods}
As students can choose to switch problem-solving strategies \emph{only} on the logic tutor, our interventions are provided in the logic training section \cite{abdelshiheed2023bridging,abdelshiheed2022power,abdelshiheed2022mixing,abdelshiheed2021preparing}. It was shown that \textit{StrTime} students frequently follow the desired behavior of switching \emph{\textbf{early}} to \textit{BC} on the logic tutor, their \textit{StrOnly} peers switch \emph{\textbf{late}}, and the \textit{Default} students make \emph{\textbf{no}} switches and stick to the default strategy \cite{abdelshiheed2022assessing,abdelshiheed2020metacognition}. Additionally, we found that providing metacognitive interventions that recommend switching to BC ---referred to as Nudges--- or present problems directly in BC ---known as Direct Presentation--- for \textit{Default} and \textit{StrOnly} students cause them to catch up with their \textit{StrTime} peers \cite{abdelshiheed2022power,abdelshiheed2022mixing,abdelshiheed2021preparing}. Therefore, we conducted two experiments to investigate different ways to present such metacognitive interventions for \textit{Default} and \textit{StrOnly} students.

\begin{figure}[ht!]
\begin{minipage}[t]{0.476\columnwidth}
\begin{minipage}[b]{1\columnwidth}
\includegraphics[width=1\linewidth]{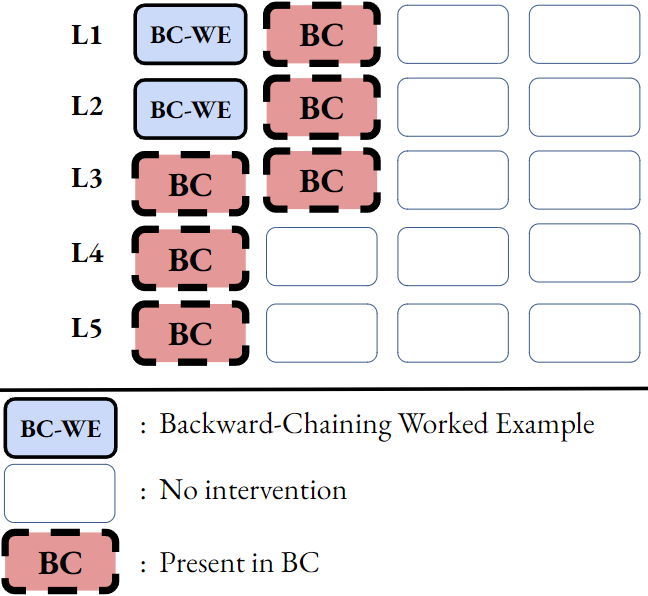}\par
\end{minipage}
\Scaption{Exp. 1: RFC$-$Static}
\label{fig:ml}
\end{minipage}\hfill
\begin{minipage}[t]{0.48\columnwidth}
\begin{minipage}[b]{\columnwidth}
\includegraphics[width=\linewidth]{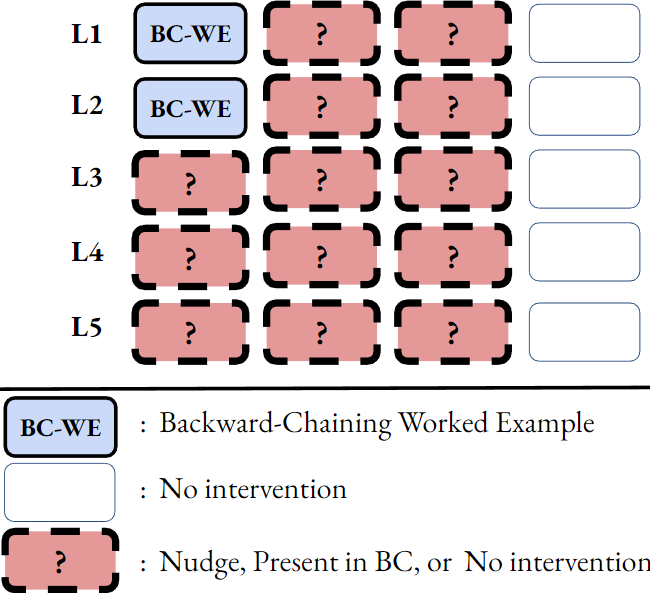}
\end{minipage}
\Scaption{Exp. 2: DRL$-$Adaptive}
\label{fig:rl}
\end{minipage}\hfill
\caption{Training on the Modified Logic Tutor} 
\label{fig:procedure}
\end{figure}

\subsection{Experiment 1: RFC$-$Static}

We utilized a RFC that could early predict a student's metacognitive group ---\textit{Default}, \textit{StrOnly} or \textit{StrTime}--- based on the incoming competence of the logic tutor. The RFC was previously shown to be $96\%$ accurate \cite{abdelshiheed2021preparing}.

After early prediction, we trained \textit{StrTime} students on the original logic tutor with all problems presented by default in forward chaining, while \textit{Default} and \textit{StrOnly} students were assigned the modified tutor, as shown in Figure \ref{fig:procedure}a. Specifically, two worked examples (WE) on BC were provided, where the tutor showed a step-by-step solution, and six problems were presented in BC by default (Direct Presentation). We expected the WEs and six problems to teach students \emph{how} and \emph{when} to use BC. Note that we selected the colored problems in Figure \ref{fig:procedure}a based on our data's historical switches to BC \cite{abdelshiheed2020metacognition}. This experiment provided a static metacognitive intervention ---Direct Presentation--- which was preferred to Nudges due to its prior success with \textit{Default} and \textit{StrOnly} students \cite{abdelshiheed2022mixing}.

\subsection{Experiment 2: DRL$-$Adaptive}

\begin{figure}[ht!]
\begin{center}
\includegraphics[width=0.72\textwidth]{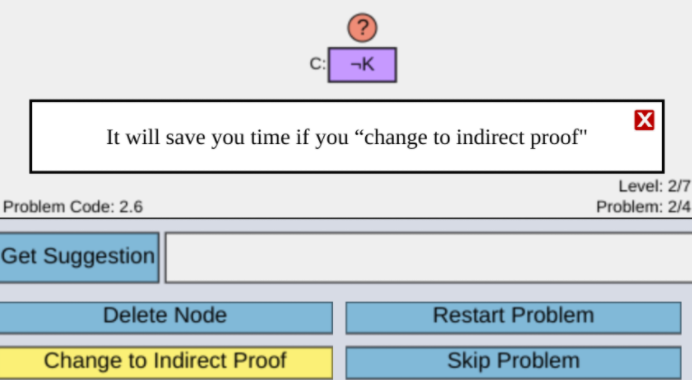}
\end{center}
\vskip -0.1 in
\caption{Strategy Switch Nudge} 
\label{fig:prompt}
\end{figure}

We leveraged DRL to provide adaptive metacognitive interventions ---\textit{Nudge}, \textit{Direct Presentation}, or \textit{No Intervention}--- regardless of the RFC's metacognitive group prediction. We trained \textit{Experimental} students on the modified tutor shown in Figure \ref{fig:procedure}b. Figure \ref{fig:prompt} shows an example of a nudge, which is prompted after a  number of seconds sampled from a probability distribution of prior students' switch behavior \cite{abdelshiheed2020metacognition}. Since our interventions included the no-intervention option, we intervened in as many problems as possible. The WEs from Experiment 1 were kept, as they are vital for teaching students \textbf{\textit{how}} to use BC. We did not intervene in the last training problem at each level, as it is used to evaluate the student's improvement on that level.

\noindent{\textbf{Training Corpus and Policy Induction:}} To train our DRL agent, we utilized data collected from four previous studies consisting of $867$ students \cite{abdelshiheed2022power,abdelshiheed2022mixing,abdelshiheed2021preparing,abdelshiheed2020metacognition} and performed a $80-20$ train-test split. The dataset consisted of a record per each student on a logic training problem represented as (\textbf{state}, \textbf{action}, \textbf{reward}). The \textit{state} is the feature vector comprising $152$ features that capture temporal, accuracy-based and hint-based behaviors. The \textit{action} is either Nudge, Direct Presentation, or No Intervention. The \textit{reward} is the immediate problem score on the logic tutor, as stated earlier in Section \ref{tutors}.

Our goal is to show that DRL works with our metacognitive interventions \textbf{rather} than \textit{which} DRL algorithm is better with our interventions. We preferred DRL to RL due to its prevailing  success in educational domains \cite{ju2021CriticalPYR,sanz2020exploring}. To select our DRL algorithm, we had to avoid a relatively simple algorithm such as Deep Q-Network (DQN), which overestimates action values \cite{mnih2015DQN} and may result in underfitting. Furthermore, we needed to avoid sophisticated DRL algorithms, such as autoencoders and actor-critic approaches, so that DRL does not overshadow the impact of our metacognitive interventions. In other words, a sophisticated DRL algorithm yielding an optimal policy would be acknowledged likely for its sophistication rather than for the metacognitive interventions it provided. Thus, we leveraged Double-DQN (DDQN), which solves the overestimation issue in DQN by \textbf{decoupling} the action \textit{selection} from the action \textit{evaluation} in two different neural networks \cite{van2016DoubleDQN}. The resulting modified Bellman equation becomes:

\begin{equation}
\label{eq3}
     Q(s, a; \boldsymbol{\theta}) = r + \gamma \, Q(s', argmax_{{a'}} Q(s', a', \boldsymbol{\theta}); \boldsymbol{\theta^-})
\end{equation}

\noindent where $r$ is the reward; $\gamma$ is the discount factor; $s$ and $s'$ refer to the current and next states; $a$ and $a'$ denote the current and next actions. Specifically, DDQN uses the \textbf{main} $(\boldsymbol{\theta})$ neural network to \textit{select} the action with the highest Q-value for the next state and then \textit{evaluates} the Q-value of that action using the \textbf{target} $(\boldsymbol{\theta^-})$ neural network. After hyperparameter tuning, we picked the model with the lowest mean squared error loss. The deployed policy had two hidden layers with $16$ neurons each, $1e$-$3$ learning late, $9e$-$1$ discount factor, $32$ batch size, a synchronization frequency of $4$ steps between main and target neural networks, and was trained until convergence ($\approx 2000$ epochs).

\section{Experiments Setup}

The two experiments took place in an undergraduate Computer Science class at North Carolina State University in the Spring and Fall of 2022, respectively. The participants were assigned each tutor as a class assignment and told that completion is required for full credit. In both experiments, students were assigned to the logic tutor following the standard procedure of pre-test, training (Fig. \ref{fig:procedure}a for Exp. 1 and Fig. \ref{fig:procedure}b for Exp. 2), and post-test. We trained students on the probability tutor six weeks later, where no interventions were provided. Each probability training problem was randomly assigned for the student to solve on their own, for the tutor to present it as a worked example (WE), or both to collaborate in the form of collaborative problem-solving. Note that on both tutors, the problem order is the same for all students.

\vskip 0.1in
\noindent{\textbf{Exp. 1 (RFC) Participants:}} A total of $121$ students finished both tutors and were classified by the RFC into $47$ \textit{Default}, $48$ \textit{StrOnly} and $26$ \textit{StrTime}. \textit{Default} and \textit{StrOnly} were randomly assigned to \textit{Experimental (RFC)} and \textit{Control (Ctrl)} conditions. \textit{Experimental} received interventions on logic training (Fig. \ref{fig:procedure}a), while \textit{Control} and \textit{StrTime} received no interventions. We had $24$ \textit{Default}$_{RFC}$, $25$ \textit{StrOnly}$_{RFC}$, $23$ \textit{Default}$_{Ctrl}$, $23$  \textit{StrOnly}$_{Ctrl}$ and $26$ \textit{StrTime} students.

\vskip 0.1in
\noindent{\textbf{Exp. 2 (DRL) Participants:}} A total of $112$ students finished both tutors and were randomly assigned to \textit{Experimental (DRL)} and \textit{Control (Ctrl)} conditions.
\textit{Experimental} received interventions on logic training (Fig. \ref{fig:procedure}b), while \textit{Control} received no interventions.  To investigate whether DRL would help students with different incoming metacognitive skills, we used the \textit{RFC to ensure even distribution across conditions and metacognitive groups for comparison purposes}. We found that our DRL policy provided no interventions for \textit{StrTime}$_{DRL}$ students $94\%$ of the time. As a result, we combined \textit{StrTime}$_{DRL}$ and \textit{StrTime}$_{Ctrl}$ into \textit{StrTime}. We had $22$ \textit{Default}$_{DRL}$, $24$ \textit{StrOnly}$_{DRL}$, $22$ \textit{Default}$_{Ctrl}$, $22$  \textit{StrOnly}$_{Ctrl}$ and $22$ \textit{StrTime} students.

\section{Results}

\subsection{Experiment 1: RFC$-$Static}

\begingroup
\renewcommand{\arraystretch}{1.3}
\begin{table}[ht!]
\scriptsize
\begin{center} 
\caption{Experiment 1 (RFC$-$Static) Results} 
\label{exp1} 
\begin{tabular}{cC{1.9cm}C{1.9cm}|C{1.9cm}C{1.9cm}|c} 
\Xhline{4\arrayrulewidth}

    & \multicolumn{2}{c|}{\textit{Experimental} (RFC)} & \multicolumn{2}{c|}{\textit{Control}} &  \\
\cline{2-5}

\makecell{}   & \makecell[t]{\textit{Default}$_{RFC}$ \\ $(N=24)$} & \makecell[t]{\textit{StrOnly}$_{RFC}$\\ $(N=25)$} & \makecell[t]{\textit{Default}$_{Ctrl}$ \\ $(N=23)$} & \makecell[t]{\textit{StrOnly}$_{Ctrl}$ \\ $(N=23)$} & \makecell[t]{\textit{StrTime} \\ $(N=26)$}  \\
\hline
\multicolumn{6}{c}{Logic Tutor}\\
\hline
\textit{Pre} &  $57.8\,(20)$ &  $56.5\,(17)$ & $55.9\,(20)$  & $55.7\,(21)$ & \cellcolor{gray!30}$57.4\,(20)$ \\
\textit{Iso. Post}  & $76.5\,(14)$ & $\mathbf{83.9\,(9)^{*}}$  & $72.3\,(16)$ & $73\,(15)$ & \cellcolor{gray!30} $81.8\,(11)^{*}$ \\
\textit{Iso. NLG} & $0.18\,(.15)$  & $\mathbf{0.32\,(.11)}^{*}$ & $0.14\,(.29)$ & $0.16\,(.32)$& \cellcolor{gray!30} $0.31\,(.17)^{*}$ \\
\textit{Post} & $73.6\,(13)$ & $\mathbf{80.5\,(9)}^{*}$  & $69.1\,(13)$ & $71.6\,(11)$ & \cellcolor{gray!30} $79.8\,(10)^{*}$ \\
\textit{NLG}  & $0.16\,(.12)$  & $\mathbf{0.31\,(.12)}^{*}$ & $0.12\,(.31)$ & $0.13\,(.3)$& \cellcolor{gray!30} $0.28\,(.16)^{*}$ \\

\hline
\multicolumn{6}{c}{Probability Tutor}\\
\hline
\textit{Pre} &  $75.5\,(16)$ & $74.2\,(15)$  &  $73.8\,(14)$ & $74.8\,(16)$ & \cellcolor{gray!30} $76.1\,(16)$ \\
\textit{Iso. Post}  & $75.2\,(17)$  & $\mathbf{93.6\,(5)}^{*}$ & $70.9\,(14)$  & $72.6\,(15)$ & \cellcolor{gray!30} $91.2\,(7)^{*}$\\
\textit{Iso. NLG} & $0.04\,(.34)$ & $\mathbf{0.35\,(.16)}^{*}$  & -$0.03\,(.29)$ & -$0.02\,(.31)$ & \cellcolor{gray!30} $0.28\,(.18)^{*}$ \\
\textit{Post} & $74.6\,(19)$  & $\mathbf{92.7\,(7)}^{*}$ & $69.5\,(16)$  & $70.9\,(17)$ & \cellcolor{gray!30} $90.3\,(8)^{*}$\\
\textit{NLG}  & $0.02\,(.37)$ & $\mathbf{0.32\,(.18)}^{*}$  & -$0.07\,(.32)$ & -$0.04\,(.36)$ & \cellcolor{gray!30} $0.25\,(.17)^{*}$ \\

\Xhline{4\arrayrulewidth}
\end{tabular} 
\end{center} 
 {\centering In a row, bold is for the highest value, and asterisk means significance over no asterisks.\par}
\end{table}
\endgroup

\noindent Table \ref{exp1} compares the groups' performance in Experiment 1. We show the mean and standard deviation of pre- and post-test scores, isomorphic scores, and the learning outcome in terms of the normalized learning gain (\textit{NLG}) \cite{abdelshiheed2020metacognition,hostetter2023self} defined as $(NLG = \frac{Post - Pre}{\sqrt{100 - Pre}})$, where $100$ is the maximum test score. We refer to pre-test, post-test and NLG scores as \textit{Pre}, \textit{Post} and \textit{NLG}, respectively. The RFC was $97\%$ accurate in classifying students who received no interventions ---\textit{Default}$_{Ctrl}$, \textit{StrOnly}$_{Ctrl}$ and $StrTime$. On each tutor, a one-way ANOVA using group as factor found no significant difference on \textit{Pre}: $\mathit{F}(4,116) = 0.21,\, \mathit{p} = .93$ for logic and $\mathit{F}(4,116) = 0.38,\, \mathit{p} = .82$ for probability.

To measure the students' improvement on isomorphic problems, repeated measures ANOVA tests were conducted (one for each group on each tutor) using \{\textit{Pre}, \textit{Iso. Post}\} as factor. We found that \textit{StrOnly}$_{RFC}$ and \textit{StrTime} learned significantly with $\mathit{p} <.0001$ on both tutors, while \textit{Default}$_{RFC}$ and the control groups did not perform significantly higher on \textit{Iso. Post} than \textit{Pre} on either tutor. These findings verify the RFC's accuracy, as \textit{StrTime} learned significantly on both tutors, while \textit{Control} did not, despite each receiving no interventions.

On both tutors, a one-way ANCOVA using \textit{Pre} as covariate and group as factor found a significant effect on \textit{Post}: $\mathit{F}(4,115) = 7.4,\, \mathit{p} < .0001, \, \mathit{\eta}^2 = 0.56$. for logic and $\mathit{F}(4,115) = 8.2,\, \mathit{p} < .0001, \, \mathit{\eta}^2 = 0.63$. for probability. Follow-up pairwise comparisons with Bonferroni adjustment $(\alpha=.05/10)$ showed that \textit{StrOnly}$_{RFC}$, \textit{StrTime} $>$ \textit{Default}$_{RFC}$, \textit{Default}$_{Ctrl}$, \textit{StrOnly}$_{Ctrl}$ on both tutors. Similar patterns were found on \textit{NLG} using ANOVA. In essence, while no significant difference was found between \textit{StrOnly}$_{RFC}$ and \textit{StrTime}, each significantly outperformed the remaining three groups.

\subsection{Experiment 2: DRL$-$Adaptive}

\begingroup
\renewcommand{\arraystretch}{1.3}
\begin{table}[ht!]
\scriptsize
\begin{center} 
\caption{Experiment 2 (DRL$-$Adaptive) Results} 
\label{exp2} 
\begin{tabular}{cC{1.9cm}C{1.9cm}|C{1.9cm}C{1.9cm}|c} 
\Xhline{4\arrayrulewidth}

    & \multicolumn{2}{c|}{\textit{Experimental} (DRL)} & \multicolumn{2}{c|}{\textit{Control}} &  \\
\cline{2-5}

\makecell{}   & \makecell[t]{\textit{Default}$_{DRL}$ \\ $(N=22)$} & \makecell[t]{\textit{StrOnly}$_{DRL}$\\ $(N=24)$} & \makecell[t]{\textit{Default}$_{Ctrl}$ \\ $(N=22)$} & \makecell[t]{\textit{StrOnly}$_{Ctrl}$ \\ $(N=22)$} & \makecell[t]{\textit{StrTime} \\ $(N=22)$}  \\
\hline
\multicolumn{6}{c}{Logic Tutor}\\
\hline
\textit{Pre} &  $55.6\,(21)$ &  $56.1\,(21)$ & $55.2\,(19)$  & $56.4\,(23)$ & \cellcolor{gray!30}$58.2\,(19)$ \\
\textit{Iso. Post}  & $91.9\,(5)^{*}$ & $\mathbf{92.1\,(4)^{*}}$  & $72.7\,(18)$ & $74.1\,(17)$ & \cellcolor{gray!30} $83.4\,(12)^{*}$ \\
\textit{Iso. NLG} & $0.46\,(.12)^{*}$  & $\mathbf{0.48\,(.09)}^{*}$ & $0.18\,(.3)$ & $0.14\,(.27)$& \cellcolor{gray!30} $0.35\,(.11)^{*}$ \\
\textit{Post} & $\mathbf{87.7\,(5)}^{*}$ & $87.6\,(5)^{*}$  & $70\,(15)$ & $69.7\,(16)$ & \cellcolor{gray!30} $80.2\,(11)^{*}$ \\
\textit{NLG}  & $\mathbf{0.45\,(.12)}^{*}$  & $0.44\,(.14)^{*}$ & $0.16\,(.33)$ & $0.1\,(.31)$& \cellcolor{gray!30} $0.31\,(.15)^{*}$ \\

\hline
\multicolumn{6}{c}{Probability Tutor}\\
\hline
\textit{Pre} &  $76.9\,(15)$ & $74.6\,(16)$  &  $75.2\,(15)$ & $76.7\,(13)$ & \cellcolor{gray!30} $78.6\,(14)$ \\
\textit{Iso. Post}  & $94.5\,(5)^{*}$  & $\mathbf{96.1\,(3)^{*}}$ & $73.9\,(10)$  & $71.4\,(14)$ & \cellcolor{gray!30} $89.1\,(7)^{*}$\\
\textit{Iso. NLG} & $0.36\,(.11)^{*}$ & $\mathbf{0.43\,(.13)}^{*}$  & -$0.02\,(.18)$ & -$0.06\,(.21)$ & \cellcolor{gray!30} $0.24\,(.15)^{*}$ \\
\textit{Post} & $94.1\,(6)^{*}$  & $\mathbf{95.6\,(4)^{*}}$ & $71.8\,(11)$  & $68.6\,(17)$ & \cellcolor{gray!30} $87.7\,(8)^{*}$\\
\textit{NLG}  &  $0.34\,(.13)^{*}$ & $\mathbf{0.39\,(.16)}^{*}$  & -$0.07\,(.24)$ & -$0.1\,(.25)$ & \cellcolor{gray!30} $0.22\,(.19)^{*}$ \\

\Xhline{4\arrayrulewidth}
\end{tabular} 
\end{center} 
 {\centering In a row, bold is for the highest value, and asterisk means significance over no asterisks.\par}
\end{table}
\endgroup

\noindent Table \ref{exp2} compares the groups' performance in Experiment 2. The RFC was $98\%$ accurate in classifying students who received no interventions ---\textit{Default}$_{Ctrl}$, \textit{StrOnly}$_{Ctrl}$ and $StrTime$. A one-way ANOVA using group as factor found no significant difference on \textit{Pre}: $\mathit{F}(4,107) = 0.18,\, \mathit{p} = .95$ for logic and $\mathit{F}(4,107) = 0.45,\, \mathit{p} = .77$ for probability. We conducted repeated measures ANOVA for each group on each tutor using \{\textit{Pre}, \textit{Iso. Post}\} as factor. We found that \textit{Default}$_{DRL}$, \textit{StrOnly}$_{DRL}$ and \textit{StrTime} learned significantly with $\mathit{p} <.0001$ on both tutors, unlike \textit{Default}$_{Ctrl}$ and \textit{StrOnly}$_{Ctrl}$.

A one-way ANCOVA using \textit{Pre} as covariate and group as factor found a significant effect on \textit{Post} on both tutors: $\mathit{F}(4,106) = 9.3,\, \mathit{p} < .0001, \, \mathit{\eta}^2 = 0.61$ for logic and $\mathit{F}(4,106) = 10.6,\, \mathit{p} < .0001, \, \mathit{\eta}^2 = 0.74$ for probability. Subsequent Bonferroni-corrected analyses $(\alpha=.05/10)$ revealed that \textit{Default}$_{DRL}$, \textit{StrOnly}$_{DRL} >$ \textit{StrTime} $>$ \textit{Default}$_{Ctrl}$, \textit{StrOnly}$_{Ctrl}$ on both tutors. For instance, \textit{Default}$_{DRL}$ had significantly higher \textit{Post} than \textit{StrTime}  ($\mathit{t}(42) = 3.9,\, \mathit{p} < .001,\, \mathit{d} = 0.88$ for logic and $\mathit{t}(42) = 3.7,\, \mathit{p} < .001,\, \mathit{d} = 0.91$ for probability) and \textit{Default}$_{Ctrl}$ ($\mathit{t}(42) = 6.6,\, \mathit{p} < .0001,\, \mathit{d} = 1.6$ for logic and $\mathit{t}(42) = 7.1,\, \mathit{p} < .0001,\, \mathit{d} = 2.5$ for probability). Similar patterns were observed using ANOVA on \textit{NLG}. In brief, the two DRL groups benefited equally from our policy, significantly outperformed their control peers, and surprisingly surpassed \textit{StrTime} students.

\vskip-0.1in

\subsection{Post-hoc Analysis}


To compare the results between the two experiments, we performed a Shapiro-Wilk normality test for each metric for each group in Tables \ref{exp1} and \ref{exp2}. The results showed no evidence of non-normality ($\mathit{p} >.05$). Therefore, we conducted independent samples t-test for every two identical groups between Tables \ref{exp1} and \ref{exp2}. Specifically, we found no significant difference between the \textit{StrTime} groups across the two tables\footnote{This holds for all metrics, such as \textit{Pre}, \textit{Post} and \textit{NLG}.}. Similarly, no such difference was observed between the \textit{Default}$_{Ctrl}$ groups or between their \textit{StrOnly}$_{Ctrl}$ peers.

The main objective was to compare our \textit{static RFC-based} and \textit{adaptive DRL-based} interventions. First, we compared the interventions' distribution within each experiment. The RFC experimental students received static interventions; hence, the distribution is identical between \textit{Default}$_{RFC}$  and \textit{StrOnly}$_{RFC}$. For DRL students, \textit{Default}$_{DRL}$ received $94 (33\%)$ Nudges, $65 (23\%)$ Direct Presentation and $127 (44\%)$ No Intervention, while \textit{StrOnly}$_{DRL}$ received $82 (26\%)$ Nudges, $74 (24\%)$ Direct Presentation and $156 (50\%)$ No Intervention. A chi-square test showed no significant difference in the distribution of interventions between the experimental DRL groups:
$\chi^2 (2,\, N=598) = 3.2, \, \mathit{p}=.2$.



Second, we compared the learning performance on both tutors. On the logic tutor, a two-way ANCOVA using \textit{Pre} as covariate, and condition \{\textit{RFC}, \textit{DRL}\} and metacognitive group \{\textit{Default}, \textit{StrOnly}\} as factors, found a significant interaction effect on \textit{Post}: $\mathit{F}(1,90) = 37.9,\, \mathit{p} < .0001, \, \mathit{\eta}^2 = 0.78$. There was also a main effect of condition: $\mathit{F}(1,90) = 28.4,\, \mathit{p} < .0001, \, \mathit{\eta}^2 = 0.59$ in that the \textit{DRL} groups significantly outperformed their \textit{RFC} peers. Follow-up Bonferroni-corrected analyses $(\alpha=.05/6)$ confirmed that \textit{Default}$_{DRL}$, \textit{StrOnly}$_{DRL} >$\textit{Default}$_{RFC}$, \textit{StrOnly}$_{RFC}$. For example, \textit{StrOnly}$_{DRL}$ had significantly higher \textit{Post} than \textit{StrOnly}$_{RFC}$: $\mathit{t}(47) = 3.6,\, \mathit{p} < .001,\, \mathit{d} = 0.98$. Similar results were found using ANOVA on \textit{NLG}.

On probability, a two-way ANCOVA using \textit{Pre} as covariate and the same two factors found a significant interaction effect on \textit{Post}: $\mathit{F}(1,90) = 29.1,\, \mathit{p} < .0001, \, \mathit{\eta}^2 = 0.46$. Subsequent pairwise analyses with Bonferroni adjustment $(\alpha=.05/6)$ revealed that \textit{Default}$_{DRL}$, \textit{StrOnly}$_{DRL}$,  \textit{StrOnly}$_{RFC} >$ \textit{Default}$_{RFC}$. For instance, \textit{StrOnly}$_{RFC}$ had significantly higher \textit{Post} than \textit{Default}$_{RFC}$: $\mathit{t}(47) = 6.4,\, \mathit{p} < .0001,\, \mathit{d} = 1.3$. Similar results were observed on \textit{NLG} using ANOVA.

\section{Discussions \& Conclusions}

We showed that leveraging DRL to provide adaptive metacognitive interventions closed the gap between metacognitive groups and caused them to surpass their control peers. Surprisingly, our DRL policy allowed the experimental students to outperform \textit{StrTime} students significantly. On the other hand, using a RFC-based approach to provide static interventions only benefited a subset of students ---\textit{StrOnly}--- who caught up with their \textit{StrTime} peers.

It is also evident that DRL prepared the experimental students for future learning \cite{bransford1999transferRethinking} by outperforming their control peers on both tutors. In other words, the experimental students outperformed their peers on probability based on interventions they received on logic. This finding suggests that they acquired backward-chaining skills in logic and transferred them to probability, where they received no interventions.

\vskip0.1in

\noindent{\textbf{Limitations and Future Work}}
There are at least two caveats in our work. First, splitting students into experimental and control conditions resulted in relatively small sample sizes. Second, the probability tutor supported only one strategy, which restricted our intervention ability to the logic tutor. The future work involves comparing the RFC-based and DRL-based approaches within one study. Additionally, we aim to make the probability tutor support forward chaining, like the logic tutor.

\vskip 0.1in

\noindent{\textbf{Acknowledgments:}} This research was supported by the NSF Grants: 1651909, 1660878, 1726550 and 2013502.

\bibliographystyle{splncs04}
\bibliography{camera_refs}

\end{document}